\documentclass[pra,aps,twocolumn,showpacs,epsfig]{revtex4}
\usepackage{amssymb}
\usepackage{amsmath}
\begin{document}

\title[Perturbative approach to the dynamics of a trapped ion]{
Perturbative approach to the dynamics of a trapped ion interacting with a light field}

\bigskip
\author{V Penna and F A Raffa}

\affiliation{
Politecnico di Torino, 
Dipartimento di Scienza Applicata e Tecnologia, Corso Duca degli Abruzzi 24, 
I-10129 Torino, Italy}

\begin{abstract}
We present a second-order perturbative analysis of the model describing 
a two-level trapped ion interacting with a traveling 
laser field, in the Lamb-Dicke regime. Unlike the customary approach, based on 
the interaction picture and the rotating wave approximation, 
we reduce the original Hamiltonian to a time-independent model, unitarily equivalent
to a spin-boson model, 
which features a form well suited to apply the perturbation theory.
We determine the first and second-order corrections of eigenstates and eigenvalues.
By varying the interaction parameters we identify four regimes characterized by
different eigenvalue distributions (possibly including level doublets) and 
suggest the presence of an intermediate regime where the spectrum undergoes macroscopic changes. 
In this regime, where the perturbation approach does not hold, our reduced Hamiltonian is shown
to reproduce the Jaynes-Cummings model. 
\end{abstract}


\maketitle


\section{Introduction}
Quantum systems formed by trapped ions interacting with a light field 
have been the subject of growing interest over the two last decades,
both due to their manifold applications in fields such as quantum measurement, 
generation of nonclassical states, and quantum-information processing 
\cite{HAFBLATT}-\cite{SK} and because they allow to engineer effective
simulators of complex quantum phenomena \cite{IIV}-\cite{JVW}.
The intense theoretical work on these systems 
and the parallel experimental progress in ion trapping are thoroughly
reviewed in \cite{HAFBLATT} and \cite{WINE}-\cite{MOYA2}. 

The distinctive feature of trapped-ion models is the time dependence introduced by
the coupling with the light field. The resulting dynamics, which is known to be
extremely complex but also very rich, has been explored by
applying or combining different well-established analytic techniques and 
approximation schemes.
The customary approach entails transforming the Hamiltonian 
of the system in the interaction picture and simplifying it through the rotating-wave 
approximation (RWA). Its study can be further simplified through the Lamb-Dicke (LD)
approximation $\ell/\lambda_L <<1$ according to which the size $\ell$ of the 
ion-confinement region is much smaller than the laser wavelength $\lambda_L$.
Applications of such an approach are well documented, e. g., in \cite{WALLS, VOGEL}, 
with reference to different ion-light resonance conditions, in the field of quantum
computation \cite{WEI,Zheng}, and in the theoretical study of the atomic population inversion 
\cite{CIRAC} and of the decoherence induced by measurement processes \cite{VIOLA}.

Non standard approach are reported as well. For instance, in \cite{WANG} and \cite{MOYA1},
by resorting to appropriate unitary transformations, the exact diagonalization of the
time-dependent Hamiltonian $H_t$ describing the ion-light field interaction is obtained also beyond 
the LD approximation, while in \cite{MOYA2} it is shown that a unitary transformation of this 
Hamiltonian leads to a mapping into the Jaynes-Cummings model, no need being to utilise the 
interaction picture and the RWA. For unrestricted values of the LD parameter this technique
holds provided the physical parameters of the system (detuning, LD parameter, trap and Rabi
frequencies) satisfy specific constraints. Interestingly, an analogous transformation scheme has
been devised in the analysis of mesoscopic spin-boson systems of trapped ions in a phonon bath \cite{porras}. 

The reduction of Hamiltonian $H_t$ to a unitarily-equivalent time-independent model
discussed in \cite{MOYA2} is also the path we follow in this work. 
We resort to appropriate unitary transformations in order to eliminate in the 
ion-light interaction term both the time dependence, controlled by the laser 
frequency $\omega_L$, and the exponential factors depending on the trapped-ion position. 
The new Hamiltonian $H_\eta$ represents a {\it minimal model},
not further reducible to a more elementary form, in which
the dependence on laser wave number $k_L$ (which is proportional
to the LD parameter $\eta$) is simply {\it linear}. It is worth noting that 
this model can be proved to be equivalent (up to a unitary transformation) to 
the well-known single-mode spin-boson model which is thoroughly reviewed in \cite{lars} 
and whose integrability has been recently analyzed in \cite{braak}.

The main aim of this paper consists in exploiting the linear character of $H_\eta$  
to diagonalize the latter, in the LD regime, {\it via} the standard perturbation method in 
which $\eta$ assumes, quite naturally, the role of perturbation parameter.
After determining the first-order corrections of the eigenstates of $H_\eta$,
we show that first-order corrections in the eigenvalue perturbation series are zero
and calculate the second-order corrections. We then focus on the resonant case where 
the laser detuning vanishes. Despite this assumption, the resulting eigenvalues feature 
a strong dependence on the interplay between the trapping-potential frequency, the 
interaction with the laser field and $\eta$. 
By varying such parameters, we identify four independent regimes corresponding to
different spectrum structures in which the eigenvalue distribution can be more or less 
dense and possibly features level doublets. The eigenstates of $H_\eta$ are shown to 
provide simple time-dependent solutions to the Schr\"odinger problem of the original
trapped-ion Hamiltonian: their superposition supplies the general solution in which 
time evolution appears to be significantly influenced by the structure of the 
eigenvalue distribution.

We also prove that applying the RWA to our Hamiltonian leads straightforwardly to
the Jaynes-Cummings model, and show that the RWA regime and the regimes where the perturbation
approach can be applied are, in a sense, complementary in that they hold in adjacent but non 
overlapping regions of the parameter space.

The layout of the paper is the following.
In section 2 we perform the reduction of the ion-light interaction model to its minimal form. 
Section 3 is devoted to the application of perturbation theory. The properties of the 
resulting spectrum are discussed in section 4 while section 5 is focused on the derivation of the Jaynes-Cummings model. Section 6 is devoted to concluding remarks.
%

\section{Minimal form of the trapped-ion model}
\label{sez2}

We consider an ion trapped in a one-dimensional harmonic potential which
interacts with a single traveling light field. The internal structure of the ion is
represented by a two-level system with ground and excited states $|g\rangle$ and $|e\rangle$,
respectively,  whose energy difference is $\hbar \omega_0$ $= $ $\hbar (\omega_e - \omega_g)$.
The relevant free Hamiltonian reads
$$
H_{\rm f} = \hbar \nu \hat{n} + \hbar \omega_0 S_z \; , 
$$
where $\nu$ is the harmonic-potential frequency, $\hat{n}$ $=$ $a^{\dagger} a$ is 
the number operator and boson operators $a$, $a^\dagger$ obey commutator $[a, a^\dagger]= 1$.
Operator $S_z =$ $ (|e\rangle\langle e|- |g\rangle\langle g|)/2$ describes the two-level 
system together with raising and lowering 
operators $S_+$ $=$ $|e\rangle\langle g|$ and $S_-= (S_+)^\dagger$ 
satisfying the standard commutators $[S_+,S_-]= 2S_z$, $[S_z,S_\pm]= \pm S_\pm$ 
of algebra su(2). 
The coupling of the ion with a light field traveling in the $x$ direction is described by
$$
H_{\textrm{int}} 
\,=\, \lambda \left[E_L \, e^{i(k_L \hat{x} - \omega_L t)}\,S_+ + {\textrm{H.c.}}\right] 
\; , 
$$
where $\lambda$ is the (real) coupling coefficient, $\hat{x}$ is the ion center-of-mass position 
operator, $E_L$ is the light (laser) amplitude, and the standard formulas 
$k_L$ $=$ $\omega_L/c$ $=$ $2 \pi/\lambda_L$ hold. 
%
By resorting to the canonical picture in terms of
displacement and momentum operators
$$
\hat{x} =  \sqrt{\hbar/(2 M \nu)} (a^\dagger + a)
\, ,\,\,\,
\hat{p} = i \sqrt{\hbar M \nu/2} \, (a^\dagger - a),
$$
the interaction Hamiltonian $H_{\rm int}$ can be expressed in an equivalent form 
depending on the LD parameter $\eta$ with $k_L \hat{x} = \eta (a^{\dagger}+a)$ and
\begin{equation}
\eta = k_L \frac{ \ell }{\sqrt 2}\, = 
{\sqrt2}\, \pi \, \frac{ \ell }{\lambda_L} \; .
\label{LDP}
\end{equation}
In formula (\ref{LDP})
$\ell =\displaystyle \sqrt{\hbar/(M \nu)}$ represents the characteristic 
length of the harmonic-confinement region and $M$ is the ion mass. 
The trapped-ion total Hamiltonian then takes the form (cf., e. g., \cite{MOYA2}, \cite{WALLS})
\begin{eqnarray}
&H_t& = H_0 + H_{\textrm{int}} \nonumber\\
&=& \hbar \nu \hat{n} + \hbar \omega_0 S_z 
+ K \left [ e^{- i \omega_L t} e^{i\eta (a^{\dagger}+a)} S_+ + {\textrm{H.C.}}
\right] , \,\,\,\,\,
\label{H}
\end{eqnarray}
where parameter $K =\lambda E_L$ has been introduced. 
%
%
\subsection{Transforming the Hamiltonian}
\label{ssez21}

The reformulation of Hamiltonian (\ref{H}) in a form suitable for the application of
perturbation theory can be realized in two steps.
First, we resort to the unitary time-dependent transformation 
$\displaystyle U_1 = e^{i \varphi S_z}$, with $\varphi$ $=$ $\varphi(t)$ $\in$ $\mathbb{R}$,
so that, by setting consistently 
$|\Psi(t)\rangle$ $=$ $ \displaystyle U_1 |\Phi(t)\rangle$, we can recast the 
Schr\"odinger problem $i \hbar \partial_t |\Psi(t)\rangle = \, H_t |\Psi(t)\rangle$ 
of Hamiltonian (\ref{H}) into the form 
$$ 
i \hbar\, \partial_t |\Phi(t)\rangle\, = \, 
\Bigl (\hbar {\dot \varphi} S_z \, + \, U_1^\dagger H_t U_1 \Bigr ) \, |\Phi(t)\rangle \; .
$$
This leads to the new (possibly time-dependent) model Hamiltonian
$H_1 = \hbar \dot{\varphi} S_z + {U_1^\dagger} H_t U_1$.
By using the customary transformation formulas 
$$
e^{i \varphi S_z} S_+ e^{-i \varphi S_z}= e^{i \varphi} S_+
\, , \,\,\,
e^{i \varphi S_z} S_- e^{-i \varphi S_z} = e^{- i \varphi} S_- \, ,
$$
representing the action of $e^{ i \varphi S_z}$ on the algebra-su(2) generators $S_{\pm}$,
and setting $\varphi = - \omega_L t$, we obtain from $H_1$ the
time-independent Hamiltonian
\begin{equation}
H_2 = \hbar \nu \hat{n} - \hbar  \Delta\, S_z 
+ K \left[e^{i \eta (a^{+}+a)} S_+ + {\textrm{ H.c.}}\right] \; ,
\label{H2}
\end{equation}
where $\Delta = \omega_L - \omega_0$ is the detuning of the system. 
It proves now convenient to perform a second
unitary transformation $\displaystyle U_2 = e^{i \beta S_z}$ with 
$\beta$ $=$ $k_L \hat{x}$ which removes 
the exponential factors depending on $ \eta \, (a^\dagger +a) \equiv k_L \hat{x}$
from $H_2$.
Observing that in ${U_2^\dagger} ( \hbar \nu \, {\hat n}) U_2 $ the term 
${\hat p}^2 $  undergoes the transformation
$$
e^{-i \beta S_z} {\hat p}^2 e^{ i \beta S_z} = \left ( \hat{p} +\hbar S_z\partial_x \beta \right)^2
$$
while harmonic term $M\nu^2 x^2/2$ is unchanged, we obtain the minimal form
for the trapped-ion model
\begin{eqnarray}
&H_\eta &  = {U_2^\dagger} H_2 U_2 
\nonumber\\
&=& \left(\hbar \nu \hat{n} + \frac{\hbar^2  k^2_L}{8M} \right) + 2 K S_x 
- \hbar \Delta S_z + \frac{\hbar k_L}{M} S_z p \, . \,\,\,
\label{H3}
\end{eqnarray}
The comparison with formula (\ref{H2}) (or (\ref{H})) 
clearly shows its considerably simpler form. 
If the eigenvalue problem $H_\eta |E(n,s) \rangle$ $= E(n,s)|E(n,s) \rangle$ is solved 
($n$ and $s$ are suitable quantum numbers relevant to the noninteracting part of $H_\eta$), 
then the general solution to the Schr\"odinger problem
$i \hbar \partial_t |\xi (t) \rangle = H_\eta |\xi (t) \rangle$
is represented by
\begin{equation}
|\xi (t) \rangle= \sum_{(n, s)} C(n,s) e^{-it E(n,s)/\hbar } |E(n,s) \rangle\, ,
\label{xisol}
\end{equation}
%
where $C(n,s)$ are arbitrary amplitudes. In this perspective,
the advantage to deal with $H_\eta$ is evident since, unlike $H_t$ or $H_2$, 
$H_\eta$ features the linear form apt to reconstruct states $|E(n,s) \rangle$
and their eigenvalues within the perturbation theory.
In this scenario ${\hbar k_L}\, S_z \, p/M$ can be interpreted as
the perturbation term since $k_L \propto \, \eta$  assumes
arbitrarily small values in the LD regime. 
%
%
\section{Perturbative scheme in the LD regime}
\label{sez3}

We rewrite Hamiltonian (\ref{H3}) in the form
$H_\eta = H_{0} + \epsilon W$ where $\epsilon W =({\hbar k_L}/{M} ) S_z p$
represents the perturbation term. The natural choice for the perturbation parameter 
is the dimensionless ratio $\epsilon = \ell/ \lambda_L$ which, up to a constant factor, 
identifies with the LD parameter $\eta$. 
Then potential $W$ is found from
\begin{equation}
W =  \frac{\hbar k_L}{\epsilon M} \, S_z p =
\, ...
= i {\sqrt 2} \pi \hbar \nu \, (a^\dagger - a) S_z .
\label{PERPOT}
\end{equation}
Formula (\ref{PERPOT}), controlled by factor $\hbar \nu$, shows that $W$ features the same energy 
scale of $\hbar \nu {\hat n}$ in the unperturbed Hamiltonian $H_0$. Then the perturbative character 
of $\epsilon W$ is ensured by the condition $\epsilon <\! < 1$. 
In passing, we observe that
the $k^2_L$-dependent term in (\ref{H3}) is not involved in the perturbative scheme
since, due to its constant character, it can be embodied in the unperturbed part $H_0$. 
According to the stationary perturbation
theory \cite{COHEN} we expand both the eigenstates and the relevant eigenvalues in powers of $\epsilon$
\begin{eqnarray}
E(n, s) &=& \sum^\infty_{k=0} \epsilon^k  E_k (n, s) \, ,\,\,\,
\label{eval}
\\
|E(n,s)\rangle &=& \sum^\infty_{k=0} \epsilon^k |E_k (n,s)\rangle\, ,
\label{evec}
\end{eqnarray}
where $s$ $=\pm$ is associated to the two-level states
$|+\rangle \equiv |e\rangle$, $|-\rangle \equiv |g\rangle$ 
such that $S_z |s\rangle$ $=$ $s |s\rangle/{2}$. With this notation one also has
$S_\pm | \pm\rangle$ $=$ 0, and
$S_s |-s \rangle$ $=$ $|s \rangle$.
%
%
%
\subsection{The unperturbed problem}
\label{ssez31}

In order to solve the unperturbed problem 
$H_0 |E_0 (n,s) \rangle$ $=$ $E_0 (n,s) |E_0 (n,s) \rangle$, we write the eigenstates 
of $H_0$ in the form
\begin{equation}
|E_0 (n,s) \rangle =  |n\rangle T^\dagger |f(s)\rangle \; , 
\label{UNP}
\end{equation}
where $|n\rangle$ is the number state such that ${\hat n}|n\rangle= n |n\rangle$, 
$T$ is a unitary transformation, and $\displaystyle T^\dagger |f(s)\rangle$ is
a spinor state defined by
\begin{equation}
T = e^{i \alpha S_y} \quad,\quad |f(s)\rangle = \frac{1}{\sqrt{2}} 
\left ( \begin{array}{c} 1 \\ s \\ \end{array} \right ) \; , 
\label{TF}
\end{equation}
%
the latter satisfying the operator equations
\begin{equation}
S_x |f(s)\rangle = \frac{s}{2} |f(s)\rangle \quad,\quad S_z|f(s)\rangle 
= \frac{1}{2} |f(-s)\rangle \; . 
\label{opeq} 
\end{equation}
The purely-spin component $2 K S_x - \hbar \Delta S_z$ of the unperturbed Hamiltonian 
%
$$
H_0  = \left(\hbar \nu \hat{n} + \frac{\hbar^2  k^2_L}{8M} \right) + 2 K S_x 
- \hbar \Delta S_z 
$$
%
is readily diagonalized by means of unitary transformation
$T^\dagger S_x T$ $=$ $S_x \cos \alpha - S_z \sin \alpha$, showing that
\begin{eqnarray}
2 K S_x - \hbar \Delta S_z &=& T^\dagger (C S_x) T \nonumber\\
&=& C(S_x \cos \alpha - S_z \sin \alpha) \; , 
\label{DIAGSPIN}
\end{eqnarray}
with the undetermined angle $\alpha$ fixed by ${\rm tg } \alpha = \hbar \Delta/(2 K)$. Then
\begin{equation}
C = \sqrt{\hbar^2 \Delta^2 + 4 K^2} \; ,\,\, \sin \alpha = \frac{\hbar \Delta}{C} \; ,\,\, 
\cos \alpha = \frac{2 K}{C} \, .
\label{alfa}
\end{equation}
%
The eigenvalues associated to $T^\dagger |f(s)\rangle$ thus coincide with those
of the problem $CS_x  |f(s)\rangle$ $=$ $ \mu_s |f(s)\rangle$. Explicitly,
$$
\mu_s = \frac{s}{2} C = \frac{s}{2} \sqrt{\hbar^2 \Delta^2 + 4 K^2}\; .
$$
Hence, the eigenvalues and the eigenstates of the unperturbed problem are given by
\begin{equation}
E_0 (n,s) = \frac{\hbar^2 k^2_L}{8M}
+ \hbar \nu \, n  + \frac{s}{2} {{\sqrt { \hbar^2 \Delta^2 + \, 4 K^2}}}  \; , 
\label{UNEIGVAL}
\end{equation}
\begin{equation}
| E_0 (n, s) \rangle = \frac{ |n \rangle }{\sqrt 2} 
\sum_{\sigma = \pm 1} s^{\frac{1-\sigma}{2}} 
\Bigl (\cos \frac{\alpha}{2} - s \sigma  \sin \frac{\alpha}{2} \Bigr ) | \sigma \rangle . 
\label{UNEIGVEC}
\end{equation}
%
%
\subsection {First-order corrections}
\label{ssez32}

In view of the linear dependence of (\ref{PERPOT}) from operators $a$ and $a^\dagger$, 
one easily shows that the first-order correction to the energy 
eigenvalues vanishes
%
$$
E_1 (n,s) = \langle E_0 (n,s) | W | E_0 (n,s) \rangle =  0 \; . 
$$
As for 
the first-order contribution to states $|E (n,s) \rangle$, 
we have
%
$$
|E_1(n,s) \rangle =   
\sum_{m\ne n} \sum_{\mu }
\frac{\langle E_0(m, \mu) | W | E_0 (n, s) \rangle}{ E_0 (n,s) -E_0 (m, \mu )} \,
|E_0(m, \mu ) \rangle \; ,
$$
in which, owing to the structure of unperturbed states (\ref{UNP}),
\begin{eqnarray}
&&
\langle E_0(m, \mu) | W | E_0 (n, s) \rangle 
\nonumber
\\
&& 
= i \sqrt{2} \pi \hbar \nu \langle m|\bigl ( a^\dagger - a \bigr ) |n\rangle \, 
\langle f(\mu)|T\, S_z T^\dagger|f(s)\rangle 
\label{NUM} 
\, . \,\,\,
\end{eqnarray}
Notice that in 
$|E_1(n,s) \rangle$ the usual specification $\mu \ne s$ in the sum with respect to 
$\mu$ is no longer required: indeed, the no-singularity condition $E_0 (m,\mu) \ne E_0 (n,s)$
is certainly verified since $\langle m | (a^\dagger - a) | n \rangle $ $=$ 0 if $m = n$, 
for both $\mu \ne s$ and  $\mu = s$. The case $\mu = s$, which should be forbidden whenever 
$m = n$, can therefore be included in the summations. 
The quantity (\ref{NUM}) can be calculated
thanks to the identities
%

%
\begin{equation}
\begin{cases}
& \langle f(s)| T\, S_z T^\dagger |f(s) \rangle = -\frac{s}{2} \sin \alpha
\cr
& {\-} \cr
& \langle f(-s)| T\, S_z T^\dagger |f(s) \rangle = \frac{1}{2} \cos \alpha
\cr
\end{cases}
\label{FMUFS} 
\end{equation}
where transformation $T S_z T^\dagger = S_z \cos \alpha - S_x \sin \alpha$
has been used. The final form of $|E_1(n,s) \rangle$ is thus found to be

$$
|E_1(n,s) \rangle
= \frac{\pi i \hbar \nu}{\sqrt{2} C}
\Bigl \{ 
\frac{s \Delta}{ \nu} \Bigl [\sqrt{n+1} | n+1,s \rangle 
+ \sqrt{n} | n-1,s \rangle \Bigr ] 
$$
\begin{equation}
+\frac{2 K \sqrt{n+1}}{-\hbar \nu + sC} 
| n+1,-s\rangle 
- \frac{2 K \sqrt{n}}{\hbar \nu + sC} |n-1,-s \rangle 
\Bigr \}  
\label{EIGVEC1}
\end{equation}
in which equations (\ref{alfa}) have been used for
$\sin \alpha$ and $\cos \alpha$, and
the simplified notation $|u, s\rangle= |E_0(n+u ,s)\rangle$ with $u = -1, 0, +1$
has been introduced.
%

\subsection{Second-order corrections and resonant case}
\label{ssez33}

Within the perturbation scheme based on states (\ref{UNP})
the expression of the second-order correction to the energy can be shown to
have the form
$$ 
E_2 (n,s) = \sum_{m\ne n} \sum_{\mu }
\, \frac{  |\langle E_0 (m, \mu) | W | E_0 (n,s) \rangle |^2   }{ E_0 (n,s) - E_0 (m,\mu)} \; .
$$
By utilizing equations (\ref{NUM}) and (\ref{FMUFS}), we find
$$
E_2 (n,s) = 2\pi^2 (\hbar \nu)^2 \left[ \sum_{\mu } 
\frac{ ( n+1)\,|\langle f(\mu)| T\, S_z T^\dagger |f(s)\rangle|^2   }{E_0 (n,s ) 
- E_0 (n+1, \mu)} \right. 
$$
$$
\left. +  \sum_{\mu } \, 
\frac{n\,|\langle f(\mu) | T\, S_z T^\dagger |f(s) \rangle|^2 }{E_0 (n,s) - E_0 (n-1,\mu)} 
\right] 
$$
\begin{equation}
= \frac{\pi^2 (\hbar \nu)^2}{2} \left[- \frac{\sin^2 \alpha}{\hbar \nu} 
+ \frac{( 2n+1) s C + \hbar \nu}{C^2 -\hbar^2 \nu^2} \cos^2 \alpha \right ], \,\,
\label{EIGVAL2}
\end{equation}
The expression of such second-order contributions remarkably simplifies
in the {\it resonant case}. Equations (\ref{UNEIGVAL}) and (\ref{EIGVAL2})
show that, in this case, 
$$
\Delta = 0 \,\,\, \Rightarrow \,\,\, \alpha = 0\, , \,\, C = 2 K\, ,
$$
leading to the second-order expression for the eigenvalues
\begin{eqnarray}
&& 
E (n,s) = E_0 (n,s) + \epsilon^2 E_2 (n,s) 
\nonumber\\
&& 
= \hbar \nu n + s K +  \pi^2 \hbar \nu \, \epsilon^2\,  
\frac{ K \bigl [ 2K  + \hbar \nu  ( 2n+1) s \bigr ]}{ 4K^2 -\hbar^2 \nu^2 } \, ,
\quad
\label{EVAL2res} 
\end{eqnarray}
where one should recall that
\begin{equation}
\epsilon^2 = \frac{\hbar k_L^2}{4\pi^2 M \nu }\, ,\qquad
K = \, \lambda\, E_L \, .
\label{pp}
\end{equation}
Note that equations (\ref{pp}) contain the significant physical parameters of the model, 
namely, the ion mass $M$, the trap frequency $\nu$, the ion-field coupling constant $\lambda$
and the laser wavenumber $k_L$ and intensity $E_L$.
In view of equations (\ref{UNEIGVEC}) and (\ref{EIGVEC1}) 
(note that for $\alpha =0$ one has $|E_0(n,s)\rangle= |n\rangle |f(s)\rangle$), 
the corresponding eigenvector proves to be
$$
|E(n,s)\rangle \simeq |E_0(n,s)\rangle + \epsilon |E_1(n,s)\rangle = 
|n\rangle |f(s)\rangle 
$$
\begin{equation} 
+ i \epsilon \frac{\pi \hbar \nu}{\sqrt{2}}  
\left[ \frac{\sqrt{n+1} | n+1,-s \rangle}{-\hbar \nu + s 2K} 
- \frac{\sqrt{n} | n-1,- s \rangle}{\hbar \nu + s 2K}  \right] \; .
\label{EVEC2res} 
\end{equation}
For the sake of completeness, the second-order correction $|E_2(n,s)\rangle$
is reported in \ref{secor}.


\section{Spectrum structure}
\label{sez4}

A significant (totally equivalent) form of eigenvalues (\ref{EVAL2res}) is 
\begin{equation}
E_* (n,s)
=
\left [ \hbar \nu  \left  (n+ \frac{1}{2} \right )+ s K \right ]
\Bigl ( 1 + s\gamma  \Bigr ) -\frac{\hbar \nu}{2}
\label{Ef}
\end{equation}
where the subscript $*$ has been introduced to evidence its factorized form
and
$$
\gamma= \frac{ 2\pi^2 \hbar \nu \epsilon^2 \, K }{4K^2 -\hbar^2 \nu^2 } \, .
$$
%

\subsection{Spectrum properties}
\label{sprop}

Expression (\ref{Ef}) is interesting because it entails an
eigenvalue distribution which, owing to $\gamma$, is strongly dependent from the 
interplay of interaction parameters. Various properties can be evinced from 
eigenvalue (\ref{Ef}).

First, it implicitly defines the parameter regions 
$$
2K < \hbar \nu\, ,\quad 2K > \hbar \nu\, ,
$$
in which $\gamma$ does not exhibit a diverging behavior,
and the value of $K$ and $\hbar \nu$ are compatible with the perturbation approach. 
In fact, the quantity ${1}/{(2K -\hbar \nu)}$ in $\gamma$ makes it evident 
how the singularities emerging when $K$ is excessively close to $\hbar \nu/2$ must be avoided.
%
Second, one easily shows that, in the limiting regimes where $2K <<\hbar \nu$ 
or $ 2K >> \hbar \nu$, $E_* (n,s)$ is characterized by 
$$
\gamma \simeq -\frac{ 2K \, \pi^2 \epsilon^2 }{\hbar \nu}
\, ,\quad 
\gamma \simeq \frac{ \hbar \nu \, \pi^2 \epsilon^2  }{2K }
\, ,
$$
satisfying, in both cases, the inequality $|\gamma| << 1$.

{\it Weak-interaction regime}. With $2K <<\hbar \nu$ one finds
$$
E_* (n,s) \simeq  -\frac{\hbar \nu}{2} \, +
\qquad \qquad \qquad\qquad\qquad
$$
\begin{equation}
\hbar \nu \left [ 
\left ( n +\frac{1}{2} \right)  
\left ( 1 -\frac{2K\pi^2\epsilon^2 s}{\hbar \nu} \right)  
+ \frac{sK}{\hbar \nu} \right ] \, ,
\label{spe1}
\end{equation}
where the fine structure of $E_* (n,s)$ is determined 
by $s K$ in which $s=\pm 1$ causes, for each $n$, the occurrence of a {\it doublet of levels}. 
In this formula the $\epsilon$-dependent term might be ignored considering that
the perturbative factor $\epsilon^2$ in $\gamma$ is further depressed
by $2K/(\hbar \nu) << 1$. This further contribution thereby represents a sort of hyperfine 
correction to the spectrum.

{\it Strong-interaction regime}.
The opposite case $ 2K >> \hbar \nu$ is more complex. One finds
\begin{equation}
E_* (n,s) 
\simeq K \left [ s +  \frac{\hbar \nu}{2K} (2n +1) + \pi^2 \epsilon^2 \frac{\hbar \nu}{2K}
\right ]  -\frac{\hbar \nu}{2} \, .
\label{spe2}
\end{equation}
The spectrum features two bands corresponding to $E_* (n', -1)$ and $E_* (n, +1)$.  
For $0 \le n' \le 2K/(\hbar \nu)$ eigenvalues $E_* (n', -1)$ are always smaller than
$E_* (0, +1)$. For $n' > 2K/(\hbar \nu)$ the levels of the band $s=-1$ intercalate those of
the band $s=+1$. In this case, a suitable choice of parameters allows the formation
of {\it interband doublets} such that $E_* (n', -1) \simeq E_* (n, +1) $ provided 
the condition 
$$
n' = n + \lfloor 2K/(\hbar \nu) \rfloor \, ,
$$
is satisfied when $2K/(\hbar \nu) = m +\delta$ is assumed where $m$ is a positive integer,
$\delta << 1$ and $\lfloor x \rfloor$ means integer part of $x$.
Then the separation between the levels of a given doublet is proportional to 
$\delta$. The fine structure of each band is represented by the $n$-dependent
term in $E_* (n,s) $ while the hyperfine structure is that corresponding to
the doublets derived from the superposition of the two bands. The perturbative factor $\epsilon^2$ 
is, in this case, further depressed by $\hbar \nu/(2K) << 1$.

{\it Intermediate regime} $\hbar \nu/(2K) \approx 1$.
A third, significant situation is found when $K$ approaches the  
value $\hbar \nu/2$ but remains far enough from it to preserve the perturbative 
character of $\gamma$. 
In this case $\gamma$ can assume both positive 
and negative values implying that, in $E_* (n, s)$, the factor $(1 +s\gamma)$ 
can decrease ($s\gamma <0$) or increase ($s\gamma >0$)
the level separation. Quantum number $s$ plays a central role in 
determining the change of the spectrum structure.

In particular, for $K \to (\hbar \nu/2)^-$ and $s= -1$ one finds
$$
s\gamma
\simeq \frac{ 2\pi^2 \, K^2 \epsilon^2 \, (-1) }{4K ( 2K -\hbar \nu) }
= + \frac{ \pi^2 \epsilon^2 \, K }{2 |2K -\hbar \nu|} \, > 0
\, ,
$$
where $|2K -\hbar \nu|$ can contrast the effect of second-order term $\epsilon^2$ 
in $\gamma$ (note that $|2K -\hbar \nu|$ represents a quantity independent from
perturbation parameter $\epsilon$ since it is fully unrelated to $\ell$ and $\lambda_L$). 
As a consequence, the separation between subsequent levels $E_* (n,-1)$ and
$E_* (n+1,-1)$ tends to grow in that factor $(1+s\gamma)$, which controls this 
separation, increases.

Instead, for $K \to (\hbar \nu/2)^-$ with $s= +1$, one finds that
$$
s\gamma
\simeq \frac{ 2\pi^2 K \epsilon^2 \, s \times 2K}{4K ( 2K -\hbar \nu) }
= - \frac{ \pi^2 \epsilon^2 \, K }{|2K -\hbar \nu|}\, < 0 \, .
$$
Again factor $\epsilon^2$ (depending on $\ell$ and $\lambda_L$) 
can be contrasted by a sufficiently small $|2K -\hbar \nu|$. 
Owing to the reduction of factor $(1 + s\gamma)$ ($\gamma <0$) the
series of eigenvalues $E_* (n,-1)$ shows a thicker level distribution.
The same effects are observed in the case $K \to (\hbar \nu/2)^+$ which displays the
decreasing (increasing) of the level separation for  $s= -1$ ($s= +1$). 
Concluding, a peculiar behavior seems to characterize the regime
$K \approx \hbar \nu/2$ suggesting that the region around 
$K = \hbar \nu/2$ (excluded from the perturbative scheme) is 
the source of a macroscopic changes of the spectrum structure. Crossing this
region evidences how, for a given quantum number $s$, a rarefied level distribution 
evolves into a thicker one while a thick distribution evolves into a more rarefied one,
depending on the direction of the crossing.
%
\subsection{Time-dependent solutions of the Schr\"odinger problem relevant to $H_t$}
\label{ssez42}

The general solution to the original Schr\"odinger problem
$i \hbar \partial_t |\Psi (t) \rangle = H_t |\Psi (t) \rangle$
is easily found to be 
\begin{eqnarray}
|\Psi (t) \rangle 
&=& U_1(t) U_2 |\xi (t) \rangle
\nonumber\\
&=& \sum_{(n, s)} C(n,s) e^{-it E(n,s)/\hbar } |\psi_t (n,s) \rangle
\label{gsol}
\end{eqnarray}
%
where $|\psi_t (n,s) \rangle \equiv U_1(t) U_2 |E(n,s) \rangle $ and
$|\xi (t) \rangle$ is defined by equation (\ref{xisol}).
$|\Psi (t) \rangle$ is the superposition of elementary time-dependent solutions
$
|\Psi_t (n,s) \rangle = e^{-it E(n,s)/\hbar }  |\psi_t (n,s) \rangle
$,
that satisfy the same Schr\"odinger problem. In the present context, 
$|\Psi_t (n,s) \rangle$
is the counterpart of the {\it stationary states} characterizing Schr\"odinger problems 
in which the Hamiltonian is time independent. Since they are in a one-to-one
correspondence with states $|E(n,s) \rangle$ they form a time-dependent basis.

The explicit form of $U_1(t) U_2 |E(n,s) \rangle$ is found by determining
the action of such transformations. Since $U_2 = \exp [{i \eta (a^\dagger +a) S_z}]$ 
one finds, to the first order in the perturbation parameter,
that state (\ref{EVEC2res})
%
(with $|n, s\rangle =|n\rangle |f(s)\rangle $ and 
$a |n\rangle$ $=$ $ {\sqrt {n}} |n-1\rangle$,
$a^\dagger |n\rangle = {\sqrt {n+1}} |n+1\rangle$) becomes
$$
U_2 |E(n,s) \rangle =
|n, s\rangle
+ 
\frac{2\pi i Ks\epsilon}{\sqrt{2}} \left[ 
\frac{a^\dagger | n, -s \rangle}{2Ks -\hbar \nu} 
- \frac{ a | n, -s \rangle}{\hbar \nu + 2Ks}  \right] \, .
$$
The action of $U_1(t) = \exp({-i\omega_L t S_z}) $ involves only spinor states $|f(s)\rangle$
giving
$$
U_1(t) |f(s)\rangle = |f(s)\rangle \cos(\omega_L t/2) -i |f(-s)\rangle \sin(\omega_L t/2)\,.
$$

An interesting aspect that deserves some comment is the considerable influence of the spectrum structure, discussed in subsection \ref{sprop}, on the general solution.
In the regime $2K << \hbar \nu$ state $ |\Psi (t)\rangle $ will have the form
$$
|\Psi (t) \rangle
= {\sum}_{n} \Bigl [ C(n,+1) e^{itK/\hbar } |\psi (n,+1) \rangle
$$
$$
+ C(n,-1) e^{-itK/\hbar } |\psi (n,-1) \rangle \Bigr ] e^{-in\nu t }\, ,
$$
showing how states $|\psi (n,\pm 1) \rangle$ of {\it level doublets} feature a slow internal 
dynamics in which frequencies $\pm K/\hbar$ could include corrections depending on 
$\epsilon^2$ (see equation (\ref{spe1})).
A different decomposition characterizes the regime $2K >> \hbar \nu$. We find
$$
|\Psi (t) \rangle
= |\Psi_S (t) \rangle + {\sum}^\infty_{n= m} \Bigl [ C(n,+1)  |\psi (n,+1) \rangle
$$
\begin{equation}
+ C(n+m, -1) e^{ i t\nu \delta } |\psi (n+m, -1) \rangle
\Bigr ] e^{-it [\nu (n+1/2) +K/\hbar]}\, ,
\nonumber
\end{equation}
where $m = \lfloor 2K/(\hbar \nu) \rfloor$, 
showing again the presence of doublet states $|\psi (n,\pm 1) \rangle$ 
the (slow) internal dynamics of which 
is controlled by parameter $\delta$ (see the comment below equation (\ref{spe2})).
State 
$$
|\Psi_S (t) \rangle = \sum^{m-1}_{n= 0} 
C(n,-1) e^{- it E(n, -1)/\hbar } |\psi (n,-1) \rangle\, ,
$$ 
involves singlet states of the low region of the spectrum
exhibiting a fast evolving dynamics.

Concerning the two intermediate regimes where $2K \simeq \hbar \nu$
(notice that $K \to \hbar\nu/2$, i. e., $K$ arbitrarily close to $\hbar\nu/2$
is forbidden) the general solution will be composed by
two macroscopic components  $|\Psi_\pm (t) \rangle$ relevant to spin states 
$s= \pm 1$ such that $|\Psi (t) \rangle = |\Psi_- (t) \rangle + |\Psi_+ (t) \rangle$. 
According to our discussion on the spectrum structure,
if the eigenvalue distribution for $s= \pm 1$ is thick then the one for $s= \mp 1$
will be rarefied.
Suppose that such an effect is emphasized when $K$ approaches 
$\hbar \nu/2$ from below. 
If the dominating components of (\ref{gsol}) involve eigenvalues $E(n, s)$   
ranging in a finite interval of ``energies" around a given average energy, 
a natural situation when a state is prepared experimentally, then
$|\Psi (t) \rangle$ will exhibit an almost continuous part corresponding to
$|\Psi_+ (t) \rangle$ superimposed to $|\Psi_- (t) \rangle$ formed by a few states 
$|\Psi_t (n, -1) \rangle$. 
Due to this very unbalanced situation we conjecture that
composition of $|\Psi (t) \rangle$ could be dramatically changed by a 
relatively small change of $K$.
This will depend on the actual size of the region around $K = \hbar \nu/2$
that separates the opposite situations where $K$ is approaching $\hbar \nu/2$ from
below and from above. In the latter case $|\Psi_- (t) \rangle$ is almost continuous 
while $|\Psi_+ (t) \rangle$ involves a few states $|\Psi_t (n, +1) \rangle$. 
The possibly macroscopic character of this effect
could be proven through an exact reconstruction of the spectrum 
around  $K = \hbar \nu/2$ {\it via} numerical calculations.

\section{The Jaynes-Cummings picture of $H_\eta$}
\label{sez5}

In this section we show that the Jaynes-Cummings (JC) model naturally emerges from 
Hamiltonian $H_{\eta }$ 
in the parameter-space region where the Rotating Wave Approximation
(RWA) is viable. This region exactly corresponds to the intermediate regime
$K \simeq \hbar \nu/2$ (and the case $K = \hbar \nu/2$ is admitted) 
where the perturbation theory cannot be implemented.
This circumstance seems to be a distinctive feature of the model described by 
Hamiltonian  (\ref{H3})
$$
H^{\rm r}_{\eta}= 
\hbar \nu \hat{n} + \frac{\hbar^2 k^2_L}{8M} \, + 2 K S_x 
+ \frac{\hbar k_L}{M}\, S_z \, p \; ,
$$
in which the resonance condition $\Delta =0$ has been assumed.
$H^{\rm r}_{\eta }$ can be expressed in an equivalent form by introducing the (nonstandard) 
raising and lowering operators 
$$
D_+ = S_y +iS_z \, , \,\,\, D_- = S_y -iS_z\, .
$$
If $S_x = D_z$ is assumed, then $D_z$, $D_-$ and $D_+$ are shown to satisfy commutators 
$$
[D_z, D_\pm] = \pm D_\pm \, ,\,\,\, [D_+, D_-] = 2D_z \, ,
$$
thereby providing an alternative (but equivalent) basis of generators 
for the algebra su(2). The link with the standard basis $S_z$ and $S_\pm$ 
is realized by the unitary transformation $U^\dagger S_\sigma U =D_\sigma$, 
$\sigma= z, \pm$, discussed in \ref{untras}. In this picture one has
\begin{equation}
H^{\rm r}_{\eta} = \hbar \nu \hat{n} + \frac{\hbar^2 k^2_L}{8M} \, + 2 K D_z 
+ \frac{\hbar k_L}{M}\, D_y \, p \, ,
\label{H3JC}
\end{equation}
with $D_y = (D_+ - D_-)/(2i) = S_z$ and $D_x =(D_+ + D_-)/2$ $=$ $S_y$.
With respect to Hamiltonian $H_\eta$, the application of the RWA implies
the condition
$$
2K \simeq \, \hbar \nu\, ,
$$ 
$K$ and $\hbar \nu$ being the characteristic energies of the non interacting part
of $H^{\rm r}_{\eta}$. In fact, it is well known, that transforming Hamiltonian 
(\ref{H3JC}) in the interaction picture, the action of operators
$D_- a$ and $D_+ a^\dagger$ is negligible since they are multiplied by fast oscillating 
factors $\exp [{\pm it (\nu  + 2K/\hbar) }]$ while $D_- a^\dagger$ and $D_+ a$ 
survive because, due to $K \simeq \hbar \nu/2$, the relevant oscillating factors 
$\exp [{\pm it (\nu  - 2K/\hbar) }]$ are essentially time independent.
Then, in
$$
D_y \, p 
= 
\frac{1}{2}  \sqrt { \frac{\hbar M\nu}{2}} \Bigl (
D_+ a^\dagger - D_+ a - D_- a^\dagger + D_- a\,  \Bigr ) \; ,
$$
the virtual processes corresponding to terms $D_- a$ and $D_+ a^\dagger$ 
are negligible, so that
%
the final form of $H^{\rm r}_{\eta}$ is
\begin{equation}
H_{JC}
=
\hbar \nu \hat{n} + \frac{\hbar^2 k^2_L}{8M} \, + 2 K D_z 
+ \Gamma \, \Bigl (D_- a^\dagger \, + D_+ a \,  \Bigr) \; ,
\label{JCH3}
\end{equation}
with $\Gamma = - { \pi \hbar \nu \epsilon }/{ \sqrt 2 } $.
%
%
Model (\ref{JCH3}) represents the JC form of the original Hamiltonian $H^{\rm r}_{\eta}$.
In this form the latter is block-diagonal: specifically, with respect to 
the states $|n-1\rangle \otimes U^\dagger|e\rangle$, $|n\rangle \otimes U^\dagger |g\rangle$, 
the $n$-th $2 \times 2$ block, $n \ge 1$, is
$$
\left[ \begin{array}{cc}
\hbar \nu (n-1) + K
&
\Gamma \sqrt{n}
\\
\Gamma \sqrt{n}
&
\hbar \nu n -K
\end{array} \right] \; .
$$
The latter reduces to $- K$ in the degenerate case $n = 0$. The nice algebraic properties of 
model (\ref{JCH3}) from which the block-diagonal structure is stemmed are reviewed 
in \ref{block}. The diagonalization of $H_{JC}$ is straightforward, the
eigenvalues being
$$
E_{JC} (n,\pm) = \hbar \nu \left(n - \frac{1}{2}\right) 
\pm \sqrt { \frac{(\hbar \nu - 2K)^2}{4} + \Gamma^2 n } \; ,
$$
where only the lower sign applies for $n = 0$. Such an expression 
appears to be very different from eigenvalues (\ref{Ef}).
When $\hbar \nu = 2K$, eigenvalues $E_{JC} (n,s)$
exhibit a linear dependence on $\Gamma \propto \epsilon$ which might be
interpreted as the signature of the transition from 
the $\hbar \nu - 2K > 0$ to the $\hbar \nu - 2K <0$ regime.

The complementary character of the JC regime, where $\hbar \nu \simeq 2K $,
with respect to regimes where $\hbar \nu > 2K $ or $\hbar \nu < 2K $
is confirmed when reformulating perturbed eigenstates 
$|E (n,s)\rangle$ in terms of operators $a$, $a^\dagger$ and $D_\pm$.
In \ref{expfo} we show that, in addition to $D_-a^\dagger$ and $D_+ a$, 
the correction $|E_1 (n,s)\rangle$ is generated by the action of $D_-a$ 
and $D_+ a^\dagger$ on $|E_0 (n,s)\rangle$, namely, by the operators 
excluded from the JC picture.

%

\section{Concluding remarks}

We have studied the time-dependent model describing the ion-light interaction
within a typical scheme which reduces the latter to a (unitarily equivalent)  
time-independent minimal form. The resulting Hamiltonian $H_\eta$ features a simple
operator structure and a linear dependence on the LD parameter $\eta$ that
have allowed us to diagonalize $H_\eta$ {\it via} the standard perturbation method.

In section \ref{sez3} we have determined the second-order analytic expressions
of the eigenstates and the eigenvalues in terms of the physically significant
parameters of the model (see equation (\ref{pp})). 
The eigenvalue formula has been used in section \ref{sez4} 
to identify four independent regimes that correspond to different spectrum structures 
in which the eigenvalue distribution can be more or less dense and possibly includes level
doublets. 
In particular, we have shown that the level density exhibits contrasting behaviors 
in the {\it intermediate} regimes characterized by $K$ ``close" to $\hbar \nu/2$: 
for $s= -1$, it is decreasing when approaching $K_0$ from below and
increasing when approaching $\hbar \nu/2$ from above. The opposite effect
characterizes the choice $s=+1$. We thus expect to observe considerable changes of
the spectrum in the unexplored neighbourhood of $\hbar \nu/2$ where the
perturbation scheme cannot be applied.
In passing, we note that similar effects have been observed in simple few-mode
bosonic models \cite{vp}. In this case the decreasing of the level separation 
has been related to the transition from stable to unstable regimes. 

In the second part of section \ref{sez4} we have reconstructed the
general solution of the Schr\"odinger problem of model $H_t$ by superposing 
the elementary solutions of minimal model $H_\eta$.
The information about the eigenvalue distribution has been used to investigate
the structure of the general solution in the four regimes characterizing the
eigenvalue spectrum.
To complete our exploration of Hamiltonian $H_{\eta}$,  
in section \ref{sez5} we have applied the RWA to $H_\eta$
reducing the latter to a Jaynes-Cummings model.
Since this approximation is based on the condition $K \simeq \hbar \nu/2$,
then the RWA leads, quite naturally, to obtain information about
the intermediate (unexplored) regime where the perturbation scheme cannot 
be applied.
In particular, we have found that the modifications of the algebraic structure 
of $H^{\rm r}_{\eta}$ caused by the RWA strongly influence the
spectrum of $H_{JC}$ changing the $\epsilon^2$-dependence into the simpler 
dependence on $\epsilon$ for $K = \hbar \nu/2$.

Future work will be focused on investigating the spectrum changes
caused by the crossing of the RWA intermediate regime. The spectral properties of the 
non-resonant case $\Delta \ne 0$ will be also explored. This requires a more extended
analysis supported by numerical calculations to get an exact description
of the observed effects. The decreasing of the level separation in certain regimes
also suggests the application of the continuous-variable picture in the 
RWA intermediate regime. This picture has been successfully applied \cite{zin,bpv}
to attractive bosonic lattices, characterized by a dense level distribution,
to study the transition between different dynamical regimes.

\acknowledgments

The financial support of the Italian government through the M.I.U.R. project
``Collective quantum phenomena: From strongly correlated systems to quantum simulators"
(PRIN 2010LLKJBX) is gratefully acknowledged.

\begin{appendix}

\section{Eigenstate second-order corrections}
\label{secor}

The second-order corrections in
$|E(n,s)\rangle \simeq |E_0(n,s)\rangle$ $+\epsilon |E_1(n,s)\rangle 
+ \epsilon^2 |E_2(n,s)\rangle$ can be easily calculated by means of
the standard formula 
$$
|E_2(n,s)\rangle = \sum_{m\ne n} \sum_r 
\frac{ \langle |E_0(m,r)|W|E_1(n,s)\rangle}{E_0(n,s)-E_0(m,r)}
|E_0(m,r)\rangle .
$$
From
\begin{eqnarray}
&W& |E_1(n,s) \rangle  = \sqrt 2 i \pi \hbar \nu (a^\dagger - a) S_z
\nonumber\\
&{\-}&\nonumber\\
&\times&
i \epsilon \frac{\pi \hbar \nu}{\sqrt{2}}  
\left[ \frac{\sqrt{n+1} | n+1,-s \rangle}{-\hbar \nu + s 2K} 
- \frac{\sqrt{n} | n-1,- s \rangle}{\hbar \nu + s 2K}  \right]
\nonumber
\end{eqnarray}
\begin{eqnarray}
&=& -\pi^2 \hbar^2 \nu^2
\Bigl [ \frac{\sqrt{n+1}\sqrt{n+2} | n+2, s \rangle}{2(2sK-\hbar \nu)} 
\nonumber\\
&{\-}& \nonumber\\
&-& \frac{\sqrt{n}\sqrt{n-1} | n-2, s \rangle}{2(2sK +\hbar \nu )}  
+ (...) | n , s \rangle  \Bigr ]\, ,
\nonumber
\end{eqnarray}
where the contribution of states $| n , s \rangle$ in $W |E_1(n,s) \rangle$
can be neglected due to the condition $m \ne n$,
one finds
$$
|E_2(n,s)\rangle = 
\pi^2 \hbar \nu
\Bigl [ \frac{ a^2| n, s \rangle}{2(2sK +\hbar \nu )} 
-\frac{ (a^\dagger)^2 | n, s \rangle}{2(2sK-\hbar \nu)} 
\Bigr ]\, .
$$
\section{New basis of su(2)}
\label{untras}

The unitary transformation $U^\dagger S_\sigma U = D_\sigma$ connects
$S_z$ and $S_\pm $ with $D_z$ and $D_\pm $, respectively. 
Its explicit definition is given by $U = \exp(i\alpha S_y)\exp(i\phi S_x)$ where 
$\alpha$ $=$ $\phi = \pi/2$. For example, with $S_- = S_x -iS_y$ one has
\begin{eqnarray}
U^\dagger S_- U
&=&
e^{-i\phi S_x} e^{-i\alpha S_y} (S_x -iS_y) S_z e^{i\alpha S_y} e^{i\phi S_x}
\nonumber\\
&=&
e^{-i\phi S_x} (S_x \cos \alpha -  S_z \sin \alpha - iS_y) e^{i\phi S_x}\, .
\nonumber
\end{eqnarray}
so that
$U^\dagger S_- U= - e^{-i\phi S_x} (S_z + iS_y) e^{i\phi S_x}$ for $\alpha= \pi/2$. 
Then the action of the $\phi$-dependent transformation gives
\begin{eqnarray}
&U^\dagger S_- U& = 
- e^{-i\phi S_x} (S_z + iS_y) e^{i\phi S_x}
\nonumber\\
&=& \!\!\! - \Bigl [ S_z \cos \phi -  S_y \sin \phi 
+i ( S_y \cos \phi +  S_z \sin \phi ) \Bigr ].
\nonumber
\end{eqnarray}
with $U^\dagger S_- U= S_y - i S_z = D_-$ for $\phi= \pi/2$.

\section{Diagonalization scheme of the JC model}
\label{block}

Hamiltonian (\ref{JCH3}) features a dynamical algebra again showing the structure
of algebra su(2). Indeed, since
$$
[D_+ a,D_- a^\dagger] = 2 D_z {N}  
\, , \, \,\, [D_+ a, {N} ] = [D_- a^\dagger, {N} ]  = 0 \; ,
$$
where ${N}$ $\doteq$ $\displaystyle (\hat{n}+  D_z + 1/2)$ 
denotes (the quantum version of) a constant of the motion relevant to 
Hamiltonian (\ref{JCH3}), one readily proves that the new
operators 
$$
J_z = D_z\, ,\,\,\, 
J_+ = \frac{1}{\sqrt {N} } D_+ a 
\, , \,\,\,
J_- = \frac{1}{\sqrt {N}} D_- a^\dagger \, ,
$$ 
satisfy $[J_+,J_-] = 2 J_z$ and $[J_z, J_\pm] = \pm J_\pm$. The latter can be seen
as the generators of a more general algebra su(2) representing the dynamical algebra 
of the JC model. In fact, up to constant term, Hamiltonian 
(\ref{JCH3}) can be expressed as a linear combination of $J_z$ and $J_\pm$
$$
H_{JC} = \hbar \nu (N-1/2) + (2K -\hbar \nu ) J_z + \Gamma \sqrt N (J_+ +J_-)\, .
$$
The algebra su(2) generated by $J_z$ and $J_\pm$ turns out to be a special case 
of the su(2) representation for generalised JC models reported in \cite{YU}.
%

\section{Operators generating the first order-correction 
$|E_1(n,s)\rangle$}
\label{expfo}
Eigenstates (\ref{EVEC2res}), including the first-order correction, can be
written as  
$$
|E(n,s)\rangle 
= |n\rangle |f(s)\rangle 
-
\frac{i \pi  \hbar \nu \epsilon  | R(n, s )\rangle }{\sqrt{2} (\hbar^2 \nu^2 - 4K^2)}
\, ,
$$
with
$$
| R(n, s )\rangle = (\hbar \nu + s 2K) a^\dagger | n, -s \rangle
+
(\hbar \nu - s 2K) a | n,-s \rangle\, .
$$
%
Owing to the identities (\ref{opeq}),
$S_x | f(s) \rangle = (s/2) | f(s) \rangle$ and 
$S_z | f(s) \rangle = (1/2) | f(-s) \rangle$,
state $|R(n,s)\rangle$ reduces to
\begin{eqnarray}
| R(n, s )\rangle
&=&
2 \hbar \nu (a^\dagger + a) S_z | n, s \rangle + 4sK (a^\dagger -a) S_z | n, s \rangle
\nonumber\\
&=&
2 \hbar \nu (a^\dagger +a) S_z | n, s \rangle +  8K (a^\dagger -a) S_z S_x | n, s \rangle
\nonumber\\
&=&
2 \hbar \nu (a^\dagger + a) S_z | n, s \rangle +  4K i (a^\dagger -a) S_y | n, s \rangle\, .
\nonumber
\end{eqnarray}
Then $| R(n, s )\rangle$ can be expressed as
\begin{eqnarray}
| R(n, s )\rangle &=&  
-i \Bigl ( (\hbar \nu -2K) ( D_+ a^\dagger  + D_- a )  \Bigr ) | n, s \rangle
\nonumber\\
&& -i \Bigl ( (\hbar \nu +2K) ( D_+ a  - D_- a^\dagger )   \Bigr ) | n, s \rangle\, ,
\nonumber
\end{eqnarray}
showing that, in addition to operators $D_+ a$ and $D_- a^\dagger$ characterizing the
JC Hamiltonian, the first-order corrections of state $|E(n,s)\rangle$
must include operators $D_- a$ and $D_+ a^\dagger$ occurring in model $H_\eta$
through the interaction term $D_y p$ (see equation (\ref{H3JC})).

\end{appendix}
%
%
%


\begin{thebibliography}{99}


\bibitem{HAFBLATT}
H\"{a}ffner H, Roos C F, Blatt R 2008 {\em Phys. Rep.} {\bf 469} 155



\bibitem{IVK} 
Ivanov S S, Vitanov N V, and Korolkova N V 2013 {\em New J. Phys.} {\bf 15} 023039
%

\bibitem{SK} 
Schmidt-Kaler F, Gulde S, Riebe M, Deuschle T, Kreuter A,
Lancaster G, Becher C, Eschner J, H{\"a}ffner H, and Blatt R
2003 {\em J. Phys. B: At. Mol. Opt. Phys.} {\bf 36} 623

\bibitem{IIV} 
Ivanov P A, Ivanov S S, Vitanov N V, Mering A, Fleischhauer M, and Singer K 
2009 {\em Phys. Rev. A} {\bf 80} 060301(R)

\bibitem{BER} 
Bermudez A, Martin-Delgado M A, and Porras D 2010 {\em New J. Phys.} {\bf 12} 123016 

\bibitem{PC} 
Porras D and Cirac J I 2004 {\em Phys. Rev. Lett.} {\bf 92} 207901 

\bibitem{JVW} 
Johanning M, Varon A F and Wunderlich C
2009 {\em J. Phys. B: At. Mol. Opt. Phys.} {\bf 42} 154009

\bibitem{WINE}
Wineland D J, Monroe C, Itano W M, Leibfried D, King B E, and Meekhof D 1998 {\em J. Res. Natl. Inst. Stand. Technol.} {\bf 103} 259
%
\bibitem{LEIB}
Leibfried D, Blatt R, Monroe C, and Wineland D  2003 {\em Rev. Mod. Phys.} {\bf 75} 281
%
\bibitem{MOYA2}
Moya-Cessa H, Soto-Eguibar F, Vargas-Mart\'{i}nez J M, Ju\'{a}rez-Amaro R, and 
Z\'{u}$\widetilde{\textrm n}$iga-Segundo A 2012 {\em Phys. Rep.} {\bf 513} 229
%
\bibitem{WALLS} 
Blockley C A and Walls D F 1993 {\em Phys. Rev. A} {\bf 47} 2115
%
\bibitem{VOGEL}
Vogel W and de Matos Filho R L 1995 {\em Phys. Rev. A} {\bf 52} 4214
%
\bibitem{WEI} 
Wei L F, Liu S Y, and Lei X L 2002 {\em Phys. Rev. A} {\bf 65} 062316
%
\bibitem{Zheng} 
Zheng X-J, Fang M-F, Liao X-P, and Cai J-W 2007 
{\em J. Phys. B: At. Mol. Opt. Phys.} {\bf 40} 507
%
\bibitem{CIRAC}
Cirac J I, Blatt R, Parkins A S, and Zoller P 1994 {\em Phys. Rev. A} {\bf 49} 1202
%
\bibitem{VIOLA}
Onofrio R, Viola L 1997 {\em Phys. Rev. A} {\bf 55} 3291
%
\bibitem{WANG}
Wang D, Hansson T, Larson \r{A}, Karlsson H O, and Larson J 2008 {\em Phys. Rev. A} {\bf 77} 053808
%
\bibitem{MOYA1}
Moya-Cessa H, Vidiella-Barranco A, Roversi J A, Freitas D S, and Dutra S M 1999 {\em Phys. Rev. A} {\bf 59} 2518
%
\bibitem{porras}
Porras D, Marquardt F, von Delft J, and Cirac J I 2008 {\em Phys. Rev. A} {\bf 78} 010101
%
\bibitem{lars}
Larson J 2007 {\em Phys. Scripta} {\bf 76} 146
%
\bibitem{braak}
Braak D, 2011 {\em Phys. Rev. Lett.} {\bf 107} 100401
%
%
%
%
\bibitem{COHEN} 
Cohen-Tannoudji C, Dui B, and Laloe F, \textit{Quantum Mechanics}, vol. II (Hermann, Paris, 1977).
%
\bibitem{YU}
Yu S, Rauch H, and Zhang Y 1995 {\em Phys. Rev. A} {\bf 52} 2585
%
\bibitem{vp}
Penna V 2013 {\em Phys. Rev. E} {\bf 87} 052909
%
\bibitem{zin}
Zin P, Chwedenczuk J, Ole B, Sacha K, and Trippenbach M 
2008 {\em Europhys. Lett.} {\bf 83} 64007
%
\bibitem{bpv}
Buonsante P, Penna V, and Vezzani A 2011 {\em Phys. Rev. A} {\bf 84} 061601


\end{thebibliography}
\end{document}